\documentclass[useAMS, usenatbib]{mn2e}
\usepackage{graphicx}
\begin{document}

\title[Fundamental parameters of RR Lyrae stars. III]
{Fundamental parameters of RR Lyrae stars from multicolour photometry
and Kurucz atmospheric models -- III. SW And, DH Peg, CU Com, DY Peg}
\author[S. Barcza and J. M. Benk\H o]
{S. Barcza
and J. M. Benk\H{o}\thanks{E-mail: barcza@konkoly.hu, 
benko@konkoly.hu}\thanks{Guest observers at Teide Observatory, Instituto de Astrofisica
de Canarias}\\
Konkoly Observatory, MTA CSFK,
		   Konkoly Thege M. \'ut 15-17.,
                   H-1121 Budapest,
                   Hungary
}
\date{Accepted 2014 May 14. Received 2014 May 14; 
in original form 2014 February 6}

\pagerange{\pageref{firstpage}--\pageref{lastpage}} \pubyear{}

\maketitle

\label{firstpage}

\begin{abstract} 
We report the most comprehensive {\it UBV(RI)$_{\mathrm C}$\/} 
observations of the bright, radially pulsating field
stars SW And, DH Peg, CU Com, DY Peg.
Long term variation has been 
found in the ultraviolet colour curves of SW And and DH Peg.
We apply our photometric-hydrodynamic method 
to determine the fundamental parameters of these stars: 
metallicity, reddening, distance, mass, radius, equilibrium 
luminosity and effective temperature.
Our method works well for SW And, CU Com and DY Peg.
A very small mass
$0.26\pm 0.04$~M$_\odot$ of SW And
has been found. The fundamental parameters of CU Com
are those of a normal double-mode RR Lyrae (RRd) star.
DY Peg has been found to have paradoxical astrophysical 
parameters: the metallicity, mass and
period are characteristic for a high-amplitude $\delta$~Sct
star while the luminosity and radius place it in the group
of RR Lyrae stars.
DH Peg has been found to be peculiar: the definite
instability in the colour curves 
towards ultraviolet, the dynamical variability of 
the atmosphere during the shocked phases suggests that
the main assumptions of our photometric-hydrodynamic method,
the quasi-static atmosphere approximation (QSAA) and the
exclusive excitation of radial modes are probably not
satisfied in this star. The fundamental parameters of all
stars studied in this series of papers are summarised in tabular 
and graphical form. 
\end{abstract}
\begin{keywords} stars: fundamental parameters 
-- stars: variables: RR Lyrae 
-- stars: individual
-- stars: atmospheres  
-- hydrodynamics.   
\end{keywords}

\section{Introduction}

The first and second parts of this series of papers (\citealt{barc4}
and \citealt{barc6},
hereafter Papers I and II) described a new 
method to determine the fundamental parameters
of spherically pulsating stars
with large amplitude, e.g. of RR Lyrae (hereafter RR) stars.
The method is purely photometric: brightness and colour
indices of ATLAS atmospheric models \citep{kuru1} are compared 
with those from multicolour observations and the obtained physical
parameters of the atmosphere are used in hydrodynamic equations for
the pulsating atmosphere. Finally, the equations are solved for 
their parameters: stellar mass 
${\cal M}_{\rm a}$
and distance 
$d$.
The parameters like the reddening
$E(B-V)$
toward the star and atmospheric metallicity
$[M/H]$
are determined from shock-free phases. The 
variable physical parameters like the effective temperature
$T_{\rm e}(t)$,
effective gravity
$g_{\rm e}(t)$,
and stellar angular radius
$\vartheta(t)=R(t)/d$
are obtained for all phases in the frame of quasi-static atmosphere
approximation (QSAA), where
$R(t)$
is the radius of zero optical depth,
$t$ 
is the time. The method
was applied for the RRab star SU Dra (Paper I) and after
some technical refinements for the double-mode (DM) stars
V500 Hya (=GSC4868-0831) and V372 Ser (Paper II).

The main trend in the research of RR
or other spherically pulsating stars with large amplitude
is nowadays to use them as distance indicators and 
calibrating their astrophysical parameters (e.g. metallicity) 
from parameters like
Fourier parameters of the light curve in an easily accessible
broad photometric band. Simple fitting formulae are sought
expressing the connection between data derived from the
photometry in one band and the astrophysical parameters
originating from involved theoretical considerations and
computations. This is essentially a statistical approach.

We emphasize the astrophysical character 
of the method presented in this series of papers.
We should like to understand 
better the response of the stellar atmosphere 
for the pulsational waves originating from the layers
deeply below the atmosphere. We do it in the frame of our
photometric-hydrodynamic method which is formulated in one
spatial dimension as well as the present theories of stellar
pulsation. We use the full colour information
of a multicolour photometry (e.g. {\it UBV(RI)$_{\mathrm C}$\/} 
in this series of papers). 
The fundamental parameters are obtained without spectroscopic 
observations. 
It is crucial to use the information content of the ultraviolet
part of the spectrum because reliable astrophysical parameters
of these stars can only be obtained if the {\it U\/}
band is included in the comparison of the observations and 
theoretical atmospheric models. 

Our method is pioneering in making use the laws 
of mass and
momentum conservation in a pulsating atmosphere. It opened a
completely new way to obtain simultaneously 
the mass and distance of the star. 

This concluding paper of the series reports 
{\it UBV(RI)$_{\mathrm C}$\/}
photometric observations and presents the new results 
for the fundamental parameters of the
stars SW And, DH Peg, CU Com, DY Peg, respectively. 

Sections 2 and 3 report the observations and the reductions. 
The results of the photometry are presented in Section 4. 
Metallicity, reddening 
derived from the colour indices of the shock-free epochs,
the variable physical parameters, mass and distance
to the stars, a brief insight in the dynamics of the atmosphere
are given in Section 5. The main results (mass, distance etc.)
from Papers I and II are also included in the tables. The 
discussion and 
conclusions are in Sections 6 and 7. These sections summarise
the results of Papers I and II as well and the appraisal of our
method. 

\section{The observations} 

A limitation in many previous photometric studies 
on these variables is that the light 
and colour curves were frequently obtained by folding observations 
over a long time 
($>10^3\times \mbox{period}$)
\citep{tifft1, liuj1} or 
one cycle was observed only \citep{paczy1, oja1}. 

We observed
segments in the Johnson-Cousins
{\it UBV(RI)$_{\mathrm C}$\/}
light curves as long as it was allowed by the sky conditions and
length of a night. Our 
light curves cover the period at least three times 
and we have the 
{\it UBV(RI)$_{\mathrm C}$\/}
magnitudes for each observed star in more than 200
epochs distributed uniformly over the
cycle. To our knowledge the photometry reported here is
the largest homogeneous observational material in the
Johnson-Cousins system containing the {\it U\/} band.
The wealth of this material allows us to discover
some hitherto unknown details of the variability,
e. g. cycle to cycle variations. 

The observations were collected with the 
IAC80\footnote{The 0.82-m IAC80 Telescope is operated on the island 
Tenerife by the Instituto de Astrofisica de Canarias in the Spanish
Observatorio del Teide.} telescope of the Teide Observatory 
and the 1-m RCC
telescope mounted at Piszk\'estet\H{o} Mountain Station of 
the Konkoly Observatory of the Hungarian Academy of Sciences. 
The observational log is given in Table~\ref{tab1}.
The exposure times were 240, 60, 40, 10, 10 s in
{\it U, B, V, R$_{\mathrm C}$, I$_{\mathrm C}$\/}
for the faintest star CU Com and
60, 40, 30, 8, 8 s for DH Peg, respectively. A selection
criterion of the target stars was that, 
according to the present day classification scheme \citep{smith1},
RRab, RRc,
RRd and SX Phe type variables should be included in our study.
Another selection criterion was that a
comparison star of similar colour and check star(s) should 
be found within the CCD frame.

\begin{table}
  \caption{Log of the observations.}
\label{tab1}
\begin{tabular}{llrl}
\hline
&HJD$-$2\,400\,000 &  No. of frames  & Telescope\\

\hline
\multicolumn{4}{l}{DY Peg, $P=0\fd072926492$} \\
&54345.3938-.7439 &  360 & IAC80  \\
&54346.4689-.7162 &  435 & IAC80  \\
&54347.4898-.7409 &  350 & IAC80  \\
\multicolumn{4}{l}{DH Peg, $P=0\fd25551037$} \\
&54349.4951-.7057 &  286 & IAC80  \\
&54352.4426-.6952 &  410 & IAC80  \\
&54354.3792-.6757 &  465 & IAC80  \\
\multicolumn{4}{l}{SW And, $P=0\fd442266$} \\
&54350.4678-.7399 &  595 & IAC80  \\
&54351.4689-.7354 &  535 & IAC80  \\
&54353.4203-.7441 &  590 & IAC80  \\
&54822.2272-.4665 &  320 & RCC  \\
&54829.2945-.4089 &  180 & RCC  \\
&54830.3834-.4302 &   75 & RCC  \\
&54831.3594-.4003 &   80 & RCC  \\
&54832.2465-.3726 &  205 & RCC  \\
\multicolumn{4}{l}{CU Com, $P_0=0\fd5439036, P_1=0\fd4056130$} \\
&54834.6136-.6923 & 50  & RCC  \\
&54871.6881-.7490$^\ast$ &  55 & IAC80\\
&54873.6045-.7612 & 135 &   IAC80\\
&54874.5278-.7652$^\ast$ & 165 &  IAC80\\
&56002.3537-.6375 & 270  & RCC  \\
&56003.3173-.5793 & 155  & RCC  \\
&56004.3384-.6196 & 235  & RCC  \\
&56005.4446-.6223 & 140  & RCC  \\
&56006.3127-.6300 & 285  & RCC  \\
&56007.3018-.6325 & 315  & RCC  \\
&56008.3011-.6224 & 310 & RCC  \\
&56018.4911-.5652 &  35 & RCC  \\
&56019.3204-.6266 & 280 & RCC  \\
&56020.3857-.6003 & 150 & RCC  \\
&56021.2945-.3516 &  60 & RCC  \\
&56022.4480-.5275 &  80 & RCC  \\
\hline
\end{tabular}
\begin{list}{}{}
\item[$^\ast$] Epoch of the tie-in observations
\item[] The source of the periods. 
        DY Peg and CU Com: this work,
        DH Peg: \citet*{rodn1},
        SW And: \citet{liuj2}.
\item[] The comparison stars were: 
GSC 1712-0984 (DY Peg),
GSC 0565-1105 (DH Peg),
GSC 1737-0809 (SW And),
GSC 1447-0968 (CU Com).
\item[] The check stars were: 
GSC 1712-1246 (DY Peg),
GSC 0565-1155 (DH Peg),
GSC 1737-1139, GSC 1737-1194 (SW And),
GSC 1447-1184 (CU Com).
\end{list}

\smallskip

\end{table}

\section{The photometric reduction}\label{sec3}

The reduction of the frames was performed in the same way described 
in Paper II. Standard {\sc iraf}\footnote{{\sc iraf} is distributed 
by the National Optical Astronomical Observatory,
operated by the Association of Universities for Research in Astronomy
Inc., under contract with the National Science Foundation.} tasks
were used and the details will not be repeated here.

The optical spectrum is sampled in
{\it U, B, V, R$_{\mathrm C}$, I$_{\mathrm C}$\/}
bands, therefore,
it is of particular importance that the photometric system
of the actual telescope and of the ATLAS models (the filters
functions, the zero points of the stellar magnitude scales) 
should be identical to avoid systematic
errors in the derived atmospheric parameters. 
Therefore, heed must be given to transforming 
the instrumental magnitudes ({\it u, b, v, r, i\/}) 
to the standard {\it UBV(RI)$_{\mathrm C}$\/} ones.
The constancy of the photometric constants of the telescope
was verified by the check stars
in order to obtain magnitudes of the best accuracy and sort
out epochs when sky conditions became insufficient to a
linear transformation between the instrumental 
and standard magnitudes.
The {\it u, b, v, r, i\/}
magnitudes were obtained in all frames
from differential photometry with respect to the comparison 
star. 

\subsection{Tie-in to standard
{\it UBV(RI)$_{\mathrm C}$\/}}\label{3.1}

\begin{table}
  \caption{The result of the photometry in the field of CU Com.}
\label{tab2}
\begin{tabular}{rrrrrr}
\hline
GSC 1447-& {\it V} &{\it B--V} & {\it U--B} & {\it V--R$_{\mathrm C}$} &{\it V--I$_{\mathrm C}$}  \\
\hline
0968     & 11.747 & 0.500& 0.042 &   0.270 &   0.560  \\
1551     & 10.889 & 0.485& 0.100 &   0.296 &   0.588  \\
1184     & 12.561 & 0.473& 0.061 &   0.261 &   0.507  \\
1247     & 14.090 & 0.781& 0.561 &   0.448 &   0.894  \\
1863     & 14.012 & 0.616& 0.237 &   0.387 &   0.732  \\
0898     & 11.586 & 0.992& 1.008 &   0.665 &   1.254  \\
\hline
\end{tabular}
\begin{list}{}{}
\item[] The errors 
of
{\it V\/},
{\it B--V\/},
etc. are
$0.004$, $0.006$, $0.013$, $0.009$, $0.009$~mag
from the 12 observations of the field. 
\item[] GSC 1447-1247 was observed by \citet{clem1} in
{\it B, V, I$_{\mathrm C}$\/},
their and our magnitudes agree within 
$1\sigma$.

\end{list}

\smallskip

\end{table}

The tie-in observations of CU Com were done under photometric
quality sky conditions. The results are summarised in Table~\ref{tab2}. 

During the observation of SW And, DH Peg, DY Peg with IAC80
the sky quality
was good only for differential photometry. We made an attempt
to tie-in observations (night
${\rm HJD}=2454348$),
however, we do not give the results because the zero points
of the magnitude scales were obviously distorted by the slightly
variable cirrus clouds over the night. To overcome this difficulty
we used the telescope constants from our previous observations 
with the telescope IAC80 on 
$\rm{HJD}-2454200=45\mbox{-}51$ \citep{barc3} to convert the
instrumental magnitude differences 
$\Delta u, \Delta b, \dots$
to international
$\Delta U, \Delta B, \dots$ ones.
Finally,
$\Delta U, \Delta B, \dots$, 
were linearly interpolated to the epoch of
{\it V\/}
observation to obtain the colour curves
for all frames reported in Table~\ref{tab1}.

To solve the problem of the zero points we folded the 
magnitudes and colours
and we shifted them to the appropriate folded observations of 
SW And \citep{liuj1} and
DY Peg \citep{kila1, oja1}
by the formula
\begin{equation}\label{3.100}
X=\Delta X+m_X^{\rm (comp)}+m_X^{\rm (z.p.)},
\end{equation}
where 
$X$
is a magnitude ($X=V, U-B, \dots$),
$m_X^{\rm (comp)}$
is the magnitude of the comparison star from the tie-in
observations on HJD=2454348 and
$m_X^{\rm (z.p.)}$
is a zero point correction. This procedure 
resulted in identical shift 
$m_X^{\rm (z.p.)}$
within $0.01$~mag for any 
$X$,
furthermore, the amplitudes and averaged magnitudes in 
{\it V\/},
{\it B--V\/},
{\it U--B\/}
became identical  
within the observational error with those of \citet{wisn1}. 

A similar procedure was applied for the observational
results of SW And with the RCC telescope.
The congruence of the light and colour curves of SW And
from the observations with the telescopes IAC80 and RCC
in the shock-free phases and the identity of 
$m_X^{\rm (z.p.)}$
with that of IAC80 at the reduction for SW And and DY Peg
indicate that our light
and colour curves are of sufficient quality to use them for
determining the atmospheric parameters of the stars. 

However, the derived shifts of zero points 
$m_X^{\rm (z.p.)}$
from SW And and DY Peg 
do not result in a congruence of light and colour curves
of DH Peg with those of \citet{tifft1}, \citet{paczy1} and
\citet{rodn1}, especially in
{\it U--B\/}.
This is caused, most probably, by cycle-to-cycle changes
in 
{\it U--B\/}
mentioned by \citet{tifft1} and \citet{wisn1}. 
This systematic variation remained hidden, because  
observations in {\it U\/} band are not available in the necessary number
(e.g. \citealt{paczy1, liuj1, oja1}).
Therefore, we fixed the zero points for DH Peg 
in a manner to reach a coincidence with the mean values
in Table~\ref{tab3}: the
corrections
$m_X^{\rm (comp)}+m_X^{\rm (z.p.)}=9.873,\: 0.283,\: 0.509,\:
0.331,\: 0.685$
were applied in Eq.~(\ref{3.100}) for
$X=V, U-B, B-V, V-R_C, V-I_C,$
respectively. 

The magnitude differences of the comparison and check stars were used in all 
fields to control the quality of the photometry
at each epoch. The standard deviation $\sigma$ of the
differences indicates the average noise of the magnitudes of the 
variables at an epoch. Of course, it is the highest for the 
faintest check star GSC 1447-1184:
$\sigma(V)=0.006$,
$\sigma(B-V)=0.007$,
$\sigma(V-R_C)=0.009$,
$\sigma(V-I_C)=0.011$,
$\sigma(U-B)=0.036$~mag are for the whole set.\footnote{
$\sigma(U-B)$ is $0.007$~mag for the 71 observations with IAC80,
respectively. This difference in
$\sigma(U-B)$
of the whole set and IAC80 subset reflects
the difference of the sky quality at the telescopes IAC80 and RCC.}

\section{Results of the photometry}

\begin{table}
  \caption{The photometric data}
  \label{tab3r}
\begin{tabular}{lllllll}
\hline
\multicolumn{7}{l}{\# SW And, folded V and colours, comp. star: GSC 1737-0968,} \\
$\cdots$ &&&&&& \\
\multicolumn{7}{l}{\#  phi, V, B-V, U-B, V-R, V-I, HJD-2400000} \\
0.0019 & 9.165 & 0.247 & 0.033 & 0.152 & 0.293 & 54350.7311 \\
0.0069 & 9.169 & 0.246 & 0.035 & 0.155 & 0.3 &	54350.7333  \\
0.0119 & 9.183 & 0.244 & 0.031 & 0.162 & 0.303 & 54350.7355  \\
$\cdots$ &&&&&& \\
\hline
\end{tabular}
\begin{list}{}{}
\item[] The complete table is published in the on-line version
as an attached file data\_FundparRRL\_III.txt.
\end{list}
\end{table}

The photometric data are published for all stars
in electronic form.\footnote{http://www.konkoly.hu/staff/benko/pub.html}

\begin{table}
  \caption{Averaged values
  of the observed stars from the $n$ epochs of the observations.}
\label{tab3}
\begin{tabular}{rrrrrr}
\hline
 & $\langle V\rangle$ & $\langle U-B\rangle$ & $\langle B-V\rangle$ 
      & $\langle V-R_C\rangle$ & $\langle V-I_C\rangle$  \\
\hline
\multicolumn{6}{l}{SW And$^1$, $n=61$, \hspace{0.06cm} 
        $\rm{HJD}-2444720=0,1,3,4$}  \\
& $9.700$ & $0.195$ & $0.423$ & $0.266$ & $0.532$ \\
\multicolumn{6}{l}{SW And, $n=344$,
       $\rm{HJD}-2454350=0,1,3$}  \\
& $9.730$ & $0.010$ & $0.496$ & $0.262$ & $0.554$ \\
\multicolumn{6}{l}{SW And, $n=172$,
       $\rm{HJD}-2454830=-8,-1,0,1,2$}  \\
& $9.746$ & $-0.024$ & $0.500$ & $0.241$ & $0.514$ \\
\multicolumn{6}{l}{SU Dra$^\dag$, $n=228$}  \\
& $9.834$ & $0.010$ & $0.311$ & $0.251$ & $0.530$ \\
\multicolumn{6}{l}{DH Peg, $n=250$}  \\
& $9.508^2$ & $0.201$ & $0.275$ & $0.190$ & $0.411$ \\
\multicolumn{6}{l}{DY Peg, $n=229$}  \\
& $10.427$ & $0.087$ & $0.294$ & $0.208$ & $0.374$ \\
\multicolumn{6}{l}{CU Com, $n=543$}  \\
& $13.313$ & $0.056$ & $0.350$ & $0.196$ & $0.427$ \\
\multicolumn{6}{l}{V372 Ser$^\ddag$, $n=529$}  \\
& $11.350$ & $0.000$ & $0.380$ & $0.256$ & $0.524$  \\
\multicolumn{6}{l}{V500 Hya$^\ddag$, $n=280$}  \\
& $10.769$ & $-0.087$ & $0.357$ & $0.215$ & $0.478$ \\
\hline
\end{tabular}
\begin{list}{}{}
\item[$^1$:] from \citet{liuj2}
\item[$^2$:] this row was calculated from the photometry 
             of \citet{tifft1}, 
\citet{paczy1}, \citet{wisn1}, and \citet{rodn1}
\item[$^\dag$:] from Paper I
\item[$^\ddag$:] from Paper II
\end{list}
\smallskip

\end{table}
The magnitude averaged 
{\it V\/}
and colour indices are in Table~\ref{tab3}.
The light and colour curves are described in the following
subsections for each star.
Before using them for a determination of the fundamental
parameters we mention some observational results 
which are interesting in themselves. 

\subsection{SW And}

Variability of the comparison star
GSC 1737-0809 (=SAO 073957) was suspected by
\citet{liuj2}. Our observations do not support it, the
magnitude differences with respect to the check stars 
GSC 1737-1194 and GSC 1737-1139 are identical within 
the observational error.
If a variability exists, its time-scale must be over 
years. We used this star as a comparison star.

\begin{figure}
  \includegraphics[width=84mm]{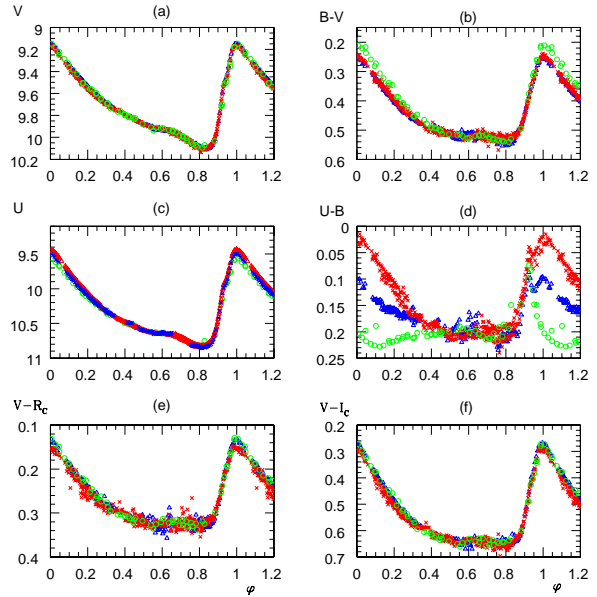}
\vspace{-0.8cm}
\caption{Light and colour curves of SW And as a 
 function of phase
 $\varphi$.
 (Green) circles: $0\le{\rm HJD}-2444720\le 4$ from the 
 photometry of
  \citet{liuj2},
 (red) crosses: $0\le{\rm HJD}-2454350\le 3$ (IAC80),
 (blue) triangles: $2\le{\rm HJD}-2454820\le 12$ (RCC).
 }
\label{fig1}
\end{figure}

We observed 62-73 percent of the full light curves on 
$\rm{HJD}-2454300=50,51,53$,
an ascending branch between
$\rm{HJD}-2454832=0.25\mbox{-}0.37$
and four shorter segments (at decreasing or minimal 
brightness). The folded light and colour curves are plotted in
Fig.~\ref{fig1}. 

The large number of our 
observations pointed out a definite variation 
about 0.04 and 0.15~mag 
in
{\it B\/}
and
{\it U\/},
respectively. 
It is clearly visible as a variation in
{\it U--B\/}
around the maximal brightness,
the hump is present in all of our observed 3 ascending branches
while it is missing in the colour curves of \citet{liuj2}. This
variation is plotted in Fig.~\ref{fig1}d.
Therefore, when fixing the zero point, our
{\it U--B\/} 
colour indices could be shifted
to the colour curve of \citet{liuj2} only in the 
shock-free phase interval
$\varphi=0.4\mbox{-}0.92$. 
An alignment is, however, impossible in the shocked intervals
$\varphi=0.93\mbox{-}1,\;0.0\mbox{-}0.4$.
A slight, much less pronounced difference is visible in the folded 
{\it B--V\/}
colour curves as well (Fig~\ref{fig1}b). The 
{\it V\/}
phase diagram and the infrared colour indices are
identical within the observational error if they are taken
from our observations and from 
\citep{liuj2}.

\citet{detre1} studied the long term behaviour of the
light curve of SW And and reported on a variable hump 
in the ascending branch and suspected a secondary (Blazhko?) 
period of $36\fd 83$. 
The hump is visible in our observations at
$\varphi\approx 0.93$
as a change of the slope in
{\it V\/}
as well as in
{\it U\/}
(Fig.~\ref{fig1}a,c).
The time coverage of our observations is not sufficient to
confirm the secondary period of \citet{detre1} because the 
same phases
$\Phi= 0.03\pm 0.05$
belong to the epochs of \citet{liuj1} and Table~\ref{tab1}
if they are folded with
$36\fd 83$. 
  
Although the amplitude variation of SW~And is
$\la 0.02$~mag 
in {\it V\/} (\citealt{barn1}, \citealt{rodn2}),
this is at the noise limit of our observations, and the 
period must be long, the 
{\it U} and {\it B\/}
observations allow us 
to confirm a change of the folded colour curves. 
The averaged colour dependence can be seen from the data in 
Table~\ref{tab3}: 
{\it B--V\/}
is redder,
{\it U--B\/}
is minimal at the maximal amplitude of the variation.
SW And follows the rule: the maximal brightness in
{\it V\/}
and the minimal
{\it U--B\/}
coincide.

\citet{liuj1} ruled out a Blazhko-type modulation
of the 
{\it V\/}
light curve during the 
$3$~days 
time-scale of their observations. Our observations confirm this
finding, however, the data in Table~\ref{tab3} and Fig.~\ref{fig1}
show clearly the long-term variation
of the light curve which is more and more pronounced toward
the ultraviolet part of the spectrum. 

\subsection{DH Peg}

\begin{figure}
  \includegraphics[width=84mm]{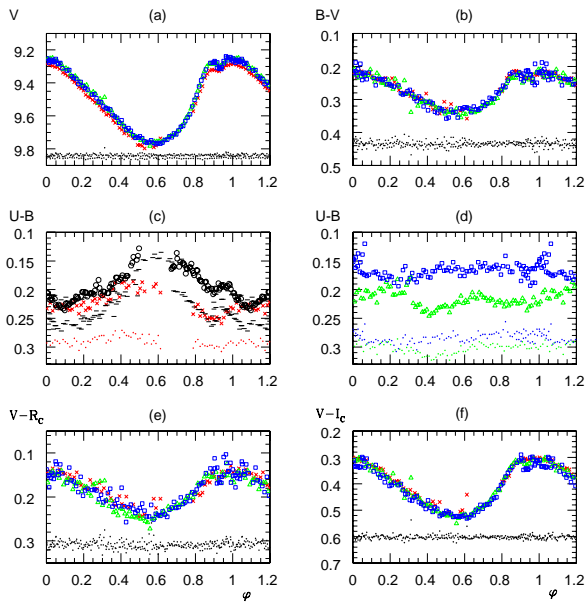}
\vspace{-0.8cm}
\caption{Folded magnitudes of DH Peg and the check star
   GSC 0565-1155.
   In all panels, (red) crosses: HJD=2454349,
   (green) triangles: HJD=2454352, 
   (blue) squares: HJD=2454354,
   dots: magnitudes of the check star GSC 0565-1155 
   calculated from the
   relative magnitudes with respect to GSC 0565-1105 and
   shifted in the panels.
   Short horizontal lines and 
   circles in panel (c): 
   {\it U--B\/} curves of \citet{tifft1} 
   and \citet{paczy1}, respectively. 
   }
\label{fig6}
\end{figure}

The phase diagrams of the check star GSC 0565-1155 and DH Peg 
are plotted in 
Fig.~\ref{fig6} with different symbols from the three nights.
The differential magnitudes 
$\Delta X$ of GSC 0565-1155 were shifted in the magnitude
range of Fig.~\ref{fig6}.
The magnitude differences of 
GSC 0565-1155 with respect to the comparison star
GSC 0565-1105 are from the 236 frames:
   $\Delta V=-0.143\pm 0.011$,
   $\Delta(B-V)=-0.435\pm 0.009$,
   $\Delta(U-B)=-0.872\pm 0.012$,
   $\Delta(V-R_C)=-0.228\pm 0.009$,
   $\Delta(V-I_C)=-0.436\pm 0.008$.
The given small standard deviations 
show the good quality of the
differential photometry within the frames. 
The following results of the differential photometry of DH Peg
are independent from the uncertainty in the zero points.

The hump before maximum is clearly visible in
all bands. 
A variation of the zero point and shape in
$\Delta(U-B)$
($\la 0.08$~mag) 
is obvious from panels \ref{fig6}c,d. It is present to 
a lesser extent in
$\Delta(B-V)(\la 0.03$~mag)
as well (Fig.~\ref{fig6}b).
This is essentially the variation of the
{\it U\/}
and
{\it B\/}
light curves amounting to
about 0.08 and 0.03~mag, respectively.
Its visibility is enhanced by looking
at the colour indices.
To demonstrate the variation in
{\it U--B\/},
the curves on
${\rm HJD}-2454300=49 \; \mbox{and} \;52,54$
were plotted separately in the panels Fig.~\ref{fig6}c and 
\ref{fig6}d, respectively.
The comparison of the three curves of DH Peg and that of
GSC 0565-1155 shows clearly that the systematic difference in
{\it U--B\/}
of DH Peg is significant above the $3\sigma$ level.    

A brightening
about 0.05~mag
of
{\it U--B\/}
is observable at
$\varphi \approx 0.55$
on $2454349$ (panel Fig.~\ref{fig6}c).
A similar, somewhat larger brightening 
($\approx 0.07$~mag)
is visible at this phase in the colour curves of \citet{tifft1} 
and \citet{paczy1}.
This brightening is missing on
$\rm{HJD}-2454400=52,54$
(Fig.~\ref{fig6}d),  
{\it U--B\/}
remained approximately constant during the whole cycle as
bright as on
$\rm{HJD}=2454349$
at
$\varphi\approx 0.55$.
A remarkable feature is that the maximal brightness in
{\it V\/}
is accompanied with maximal or approximately constant  
{\it U--B\/},  
this is visible in the material of
\citet{tifft1} and \citet{paczy1}, as well.

The three observed ascending branches in the near infrared 
colours are identical within the scatter, while a minor
systematic difference ($\la 0.03$~mag) is observable in the 
descending branches. 
Because of the uncertainty of zero points
of the magnitudes, our amplitudes were compared with those
of the published previous studies. The near infrared amplitudes
are identical with those from \citet{rodn1}. The amplitudes
in 
{\it B--V\/}, {\it U--B\/}
show differences
$\approx 0.02$ and 0.06~mag,
respectively. 

\subsection{CU Com}

\begin{figure}
  \includegraphics[width=84mm]{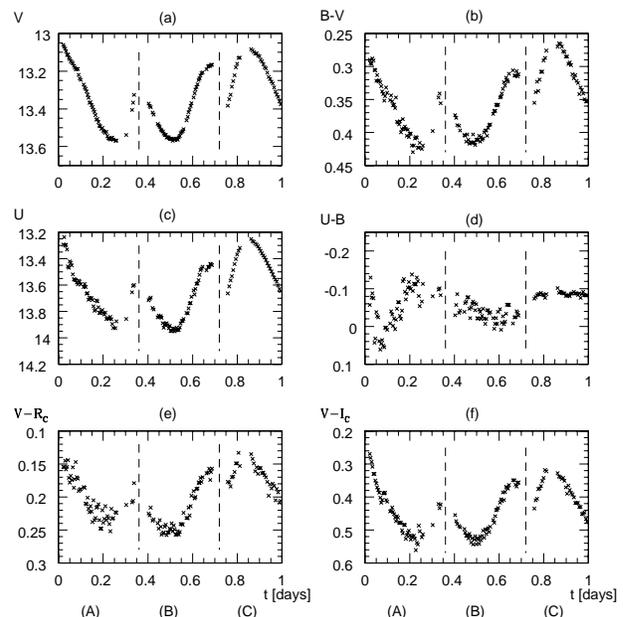}
\vspace{-0.8cm}
\caption{Three characteristic segments from the photometry of CU Com.
To unify the segments in one figure, the following time shifts of 
the observed epochs were applied:
$t_{\rm (A),(B),(C)}=\mbox{HJD}-2456008.28\; \mbox{(A)}$, 
$-2456001.95\;\mbox{(B)}$ 
and
$-2454873.77\;\mbox{(C)}$,
respectively.}
\label{fig5}
\end{figure}

The folded light curves are not informative because
of the presence of two periods with approximately same amplitude,
therefore, three characteristic segments are plotted in
Fig.~\ref{fig5}: a descending branch (A), a minimum (B)
and a maximum (C). It is remarkable that
{\it U\/}
has a definite hump before maximum in segment (B), Fig.~\ref{fig5}c 
and the slope of
{\it V\/}
shows a change here (Fig.~\ref{fig5}a). This feature
is present in the 7 segments containing an ascending branch 
on
$\mbox{HJD}-2456000=2, 3, 4, 6, 7, 19, 20$. 
CU Com follows the rule: maximal brightness in
{\it V\/}
coincides with the minimal
{\it U--B\/}.

A Fourier analysis of the {\it V\/} light curve
by the {\sc mufran} program package \citep{Muf} 
yielded the periods
$P_0=f_0^{-1}=0\fd 5439036\pm 0\fd0000030$
$P_1=f_1^{-1}=0\fd 4056130\pm 0\fd0000017$,
the period ratio is
$P_1/P_0=0.745744$.
The rest mean square of the residual is
$0.016$~mag 
after prewhitening the frequencies
$if_0,\; i=1,2,3,4$,
$jf_1,\; j=1,2,3$
and
$f_0\pm f_1$
from the light curve. The frequencies
$f_0\pm f_1$
could be definitely identified, that is, 
typical frequencies for an RRd star have been found. Small 
differences 
$\Delta P_0=-0\fd00026$
and
$\Delta P_1=-0\fd00015$
of the periods can be seen in
comparison with those from \cite{clem1}.
This is equivalent to secular period changes
${\dot P_0}\approx -4.8\times 10^{-8}$
and
${\dot P_1}\approx -2.7\times 10^{-8}\;\mbox{days}^{-1}\mbox{days}$
which are by a factor
$\mbox{O}(10^3)$
larger than the values expected from stellar evolution theory and
$\dot P_0$,
$\dot P_1$
found for the RRd star V372 Ser \citep{barc3}.
The period ratio
$P_1/P_0=0.745658$
\citep{clem1} changed 
$\approx +1.15\times 10^{-4}$
over some
$15$~years.

\subsection{DY Peg}\label{sec4.4}

\begin{figure}
  \includegraphics[width=84mm]{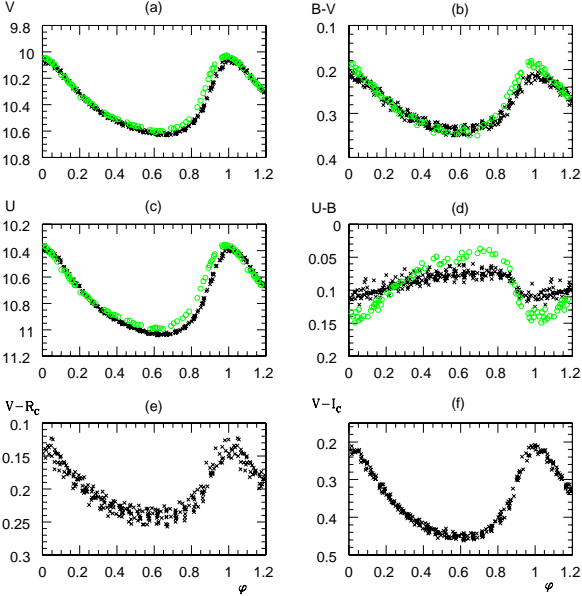}
\vspace{-0.8cm}
\caption{Folded light and colour curves of DY Peg.
 (Green) circles: from the 
 {\it UBV\/}
 photometry of \citet{oja1} (${\rm HJD}-2450387=0.34\mbox{-}0.49$),
 black crosses: from the IAC80 observations on 
 ${\rm HJD}-2454340=5, 6, 7$.
  }
\label{fig4}
\end{figure}

The light and colour curves of DY Peg are plotted in 
Fig~\ref{fig4}. 
An evident feature 
is that the maximal brightness in
{\it V\/}
is accompanied with a maximal   
{\it U--B\/}.
DY Peg behaves contrary to an RRab star having maximal
{\it V\/}
and minimal
{\it U--B\/}
simultaneously. A similar behaviour was observed at DH Peg. 
The other colour indices of DY Peg and RR stars
show identical qualitative feature: the maximal brightness in
{\it V\/}
is accompanied with maximal brightness in
{\it B--V\/}, 
{\it V--R$_{\mathrm C}$\/} and
{\it V--I$_{\mathrm C}$\/}.

We measured some 0.03~mag smaller 
amplitude of
{\it U--B\/} 
in comparison with that of Oja (2011). This might be connected
with the amplitude variation of the {\it V\/} light curve 
which has already been reported by numerous 
authors (e.g. \citealt*{Koz80, GR96, Pop03, Fu09})
who explained it by multiperiodic pulsation. A variability 
$\approx 0.04$~mag 
was observed in 
{\it V\/} between two complete light curves taken with 12 days 
difference by \citet{meyla1}.

The main pulsation frequency 
$f_0= 13.712438$~d$^{-1}$ 
($P=0\fd 072926492$, see Table~\ref{tab1})
and its four significant
harmonics were detected by the {\sc mufran} program 
package.
After these frequencies were pre-whitened
from the data, the spectrum of the residual still has some 
structure. 
A wide peak can be found at around 17.67~d$^{-1}$ with the
amplitude of $\approx 4$~mmag. The resolution of the
spectrum is very limited because of our short observing run,
therefore, this peak is not significant ($2\sigma)$. 
However, the position and amplitude of this peak agrees well with
the frequency of the first overtone pulsation reported 
in the cited literature. The short time span of our observations
does not allow us to draw more quantitative conclusion on
significant other period(s) and amplitudes belonging to them.

\section{Astrophysical parameters}

The astrophysical parameters of the target stars were
determined using the {\sc bbk} package\footnote{The package
is available in electronic form: 
{\tt http://www.konkoly.hu/staff/barcza.shtml/publications}} \citep{barc7} as follows.

The metallicity 
$[M/H]$
and the reddening
$E(B-V)$
toward the stars were determined 
by the minimization of the averaged
errors of the effective temperature 
$\overline{\Delta T_{\rm e}}$
and gravity 
$\overline{\Delta \log g_{\rm e}}$
from the possible
combinations of the colour indices
\citep{barc8} using the photometry in the shock-free
epochs. The upper limit of the search for
$E(B-V)$ 
was taken from the maps of the satellite Diffuse 
Interstellar Background Explorer ({\it DIRBE}, \citealt*{schl1}).
The results are given in Table~\ref{tab4}.

\begin{table}
  \caption{Metallicity, reddening, averaged surface gravity, effective
           temperature, angular radius of the stars and the
	   standard deviation of the averaged quantities.
	   }
\label{tab4}
\begin{tabular}{lllll}
\hline
 $[M/H]$  & $E(B-V)$ & $\overline{\log g_{\rm e}}$ 
        & $\overline{T_{\rm e}}$ 
	& $\overline{\vartheta\times 10^{11}}$ \\
 $\mbox{[dex]}$ & $\mbox{[mag]}$ & [cms$^{-2}$] 
	& [K] & [radians] 	\\
\hline
\multicolumn{5}{l}{SW And, 
        $\rm{HJD}-2444720=0,1,2,3,4$}  \\
        & $0.02$ & $2.60\pm .35$ & $6676\pm 421$ 
        & $17.52\pm .68$  \\
\multicolumn{5}{l}{\hspace{1.22cm}
        $\rm{HJD}-2454350=0,1,3$}  \\
 $0.10$  & $0.02$ & $2.85\pm .63$ & $6610\pm 428$ 
        & $17.31\pm .68$ \\
\multicolumn{5}{l}{\hspace{1.22cm} 
        $\rm{HJD}-2454830=-8,-1,0,1,2$}  \\
        & & $2.75\pm .44$ & $6637\pm 393$ 
        & $17.53\pm .65$  \\
\multicolumn{5}{l}{SU Dra$^\dag$} \\
 $-1.60$ &$0.015$ & $2.72\pm .59$ & $6743\pm 512$ 
        & $16.68\pm .91$  \\
\multicolumn{5}{l}{DH Peg} \\
 $ -0.35$ & $0.08^\ast$ & $2.86\pm .43$ & $7413\pm 312$  
        & $16.90\pm .04$  \\
\multicolumn{5}{l}{DY Peg} \\
 $-0.05$ & $0.0$  & $3.41\pm .16$ & $7157\pm 246$ 
        & $10.60\pm .19$ \\
\multicolumn{5}{l}{CU Com} \\
 $-2.20$ &$0.02$ & $3.09\pm .51$ & $6925\pm 309$ 
        & $3.21\pm .12$ \\
\multicolumn{5}{l}{V372 Ser$^\ddag$} \\
 $-0.53$ &$0.003$& $3.24\pm .36$ & $6713\pm 324$ 
        & $8.13\pm .16$ \\
\multicolumn{5}{l}{V500 Hya$^\ddag$} \\
 $-1.05$ & $0.008$& $3.69\pm .50$ & $6902\pm 261$ 
        & $10.17\pm .37$ \\
\hline
\end{tabular}

\smallskip
\begin{list}{}{}
\item[] 
The following phase intervals were used in the folded colour 
curves to derive
$[M/H]$ 
and
$E(B-V)$.

SW And: $0.4 < \varphi < 0.8$, (${\rm HJD}-2454300=50,51,53$), 

DH Peg: $0.1 < \varphi < 0.4$, (${\rm HJD}= 2454349$),

DY Peg: $0.3 < \varphi < 0.8$, (${\rm HJD}= 2454300=45,46,47$),

CU Com: 39 epochs were taken from the shock-free 
intervals on
${\rm HJD}-2454800=71,73,74$.

The errors are $\pm 0.10\:\mbox{dex},\:\pm 0.01\:\mbox{mag}$,
(except for CU Com where it is $\pm 0.20\:\mbox{dex}$),
respectively.
\item[$^\ast$] \citet{rodn1}
\item[$\dag$] Paper I
\item[$\ddag$] Paper II
\end{list}

\end{table}

The colour-colour diagrams of the ATLAS
models \citep{kuru1} were interpolated to the values of
$[M/H]$
and 
$E(B-V)$
in Table~\ref{tab4} and they were used 
to determine 
$\vartheta$,
$T_{\rm e}$
and
$g_{\rm e}$
as a function of phase. Their averaged values from our
observations are given also in Table~\ref{tab4}. 

Next, the time dependent quantities were introduced
in the hydrodynamic equations which were solved for 
$d$
and
${\cal M}_{\rm a}$ 
as described in Paper II. 
One point was added to the algorithm {\sc bbk}: the lower limit 
$d_{\rm min}=-g_{\rm e}(\varphi_1)
  {\ddot\vartheta}^{-1}(\varphi_1)$
of the search was introduced to exclude false roots of 
Eq. (4) of Paper II coming from terms
$\propto d^{-1}$,
$\varphi_1$
is the phase of minimal
$g_{\rm e}$,
$\ddot\vartheta$
is the angular acceleration at the top of the atmosphere
(in the reference frame of the observer), the dot denotes a 
differentiation with respect to
$t$.
The results are summarised in Table~\ref{tab5}.
The following variable atmospheric parameters
are plotted for each star in Figs.~\ref{fig2}-\ref{fig9}:
angular radius $\vartheta(\varphi)$, effective gravity
$g_{\rm e}(\varphi)$, effective temperature $T_{\rm e}(\varphi)$,
barometric scale height for unit averaged molecular mass $\mu$ at
the top of the atmosphere  
$\mu h^{-1}_0(\varphi)={\cal R}T(R,\varphi)g^{-1}_{\rm e}$,
radius $R(\varphi)$ and pulsational velocity $v(\varphi)$,
$\cal R$
is the universal gas constant.

\begin{table*}
\begin{minipage}{170mm}
  \caption{
  ${\cal M}_{\rm a}d^{-2}$, distance
  $d$
  of the observed stars, the averaged residual acceleration
  of the atmosphere in the epochs when the dynamical condition
  $\rm{C^{(II)}}$ (Paper I) of QSAA is satisfied, 
  mass, equilibrium 
  luminosity and
  effective temperature, minimal and maximal radius, 
  magnitude averaged absolute visual magnitude, averaged
  `static' surface gravity.}
\label{tab5}
\begin{tabular}{llllllllll}
\hline
  & ${\cal M}_{\rm a}d^{-2}\times 10^7$ & $d$ &
  $\overline{a^{\rm (dyn)}}$ 
        & ${\cal M}_{\rm a}$ & $L_{\rm eq}$ & $T_{\rm eq}$ 
	& $R_{\rm min},R_{\rm max}$ 
	& $\langle M_V \rangle$ & $g_{\rm s}$ \\
	& [M$_\odot\mbox{pc}^{-2}$] & [pc]   
        & [$\mbox{ms}^{-2}$] & [M$_\odot$] 
	& [L$_\odot$] & [K] & [R$_\odot$] & [mag]
	& [$\mbox{ms}^{-2}$] \\
\hline
$\mbox{SW And}^\ast$ 
         & $7.00\pm 0.76$ & $626\pm 31$ & $\phantom{-}0.27$ &
         $0.26\pm 0.04$ & $39.8\pm 4$ & $6644$ &
         $4.51,5.05$ & $0.710$ & $3.09$\\
$\mbox{SW And}^{\ast\ast}$ 
         & & & $\phantom{-}1.685$ &
         & $41.5\pm 4$ & $6672$ &
         $4.53,5.06$ & $0.68$ & $3.02$\\
$\mbox{SW And}^{\ast\ast\ast}$ 
         &  &  &  & 
         & $41.8\pm 4$ & $6690$ &
         $4.44,5.05$ & $0.655$ & \\
SU Dra$\dag$ 
         & & $663\pm 67$ & & $0.68\pm 0.03$
         & $45.9\pm 9.3$ & $6813$ 
	 & $4.46,5.29$ & $0.68$ & $8.3$\\ 
DH Peg$^\bullet$   
         & $9.20\pm 0.32$ & $893\pm 104$ & $-0.03$ 
         & $0.73\pm 0.09$ & $123\pm 27$& $7464$ & $6.40,7.06$
	 & $0.02$ & $4.50$ \\
DY Peg   & $2.42\pm 0.27$ & $817\pm 24$ & $\phantom{-}1.40$ 
         & $1.40\pm 0.24$ & $34.6\pm 2.1$ & $7177$  
	 & $3.74,3.95$ & $0.84$ & $30.2$ \\
CU Com   & $1.28\pm 0.21$ & $3059\pm 181$ & $\phantom{-}0.02$
         & $0.55\pm 0.03$ & $39.0\pm 4.7$ & $6942$ 
	 & $3.95,4.70$ & $0.82$ & $8.00$ \\ 
V372 Ser$^\ddag$
         & $6.12\pm 0.31$ & $964\pm 81$ & $\phantom{-}0.41$ 
	 & $0.57\pm 0.10$ & $21.9\pm 5.2$ & $6722$ 
	 & $4.07,4.40$ & $1.58$ & $12.9$ \\
V500 Hya$^\ddag$
         & $40.3\pm 6.7$ & $467\pm 16$ & $\phantom{-}2.40$ 
	 & $0.88\pm 0.06$ & $8.97\pm 1.23$ & $6924$ 
	 & $1.97,2.05$ & $2.40$ & $54.8$ \\
\hline
\end{tabular}
\begin{list}{}{}
\item[] The estimated errors of
      $[M/H]$, $T_{\rm eq}$, $\langle M_V \rangle$ are
      $\pm 0.1\mbox{dex}, 25\mbox{~K}, 0.25\mbox{~mag}$,
      respectively,  
      $g_{\rm s}=G{\cal M}\overline{R}^{-2}$
\item[$^\ast$] HJD$-2454350=0,1,3$
\item[$^{\ast\ast}$] HJD$-2454830=-8,-1,0,1,2$, ${\cal M}_{\rm a}$, $d$
     of $^\ast$ were used
\item[$^{\ast\ast\ast}$] HJD$-2444720=0,1,3,4$, ${\cal M}_{\rm a}$, $d$
     of $^\ast$ were used
\item[$\dag$] Paper I
\item[$\ddag$] Paper II
\item[$\bullet$] The data were determined
      from the photometry on $\rm HJD=2454349$, see text.
\end{list}

\end{minipage}

\smallskip

\end{table*}

The photometric condition
$\rm C^{(I)}$
of the QSAA (Paper I) is not satisfied at the phase of maximal
compression of the atmosphere in any of the stars. Therefore,
some (positive) correction to
$g_{\rm e}(\varphi), T_{\rm e}(\varphi),$
can be expected here, however, the qualitative features of the 
curves remain unchanged: their maximal values are at the maximal
luminosity and brightness in
{\it V\/}.
This correction does not have an effect on
${\cal M}_{\rm a}$
and
$d$
because they are determined from shock-free phases. Nevertheless,
an effect on
$\vartheta(t)$
and, of course, on
${\dot\vartheta}$,
and
${\ddot\vartheta}$
can be expected.

\subsection{SW And}

The metallicity
$[M/H]=+0.1\pm 0.1$~dex
from our photometric minimization method 
and
$T_{\rm eq}=6644$~K agree well with
$[M/H]$
and
$T_{\rm e}$
derived from high-dispersion spectra \citep{neme1}.
Remarkable is the small mass
${\cal M}_{\rm a}=0.26\pm 0.04$~M$_\odot$
(Table~\ref{tab5}).  

\begin{figure}
  \includegraphics[width=84mm]{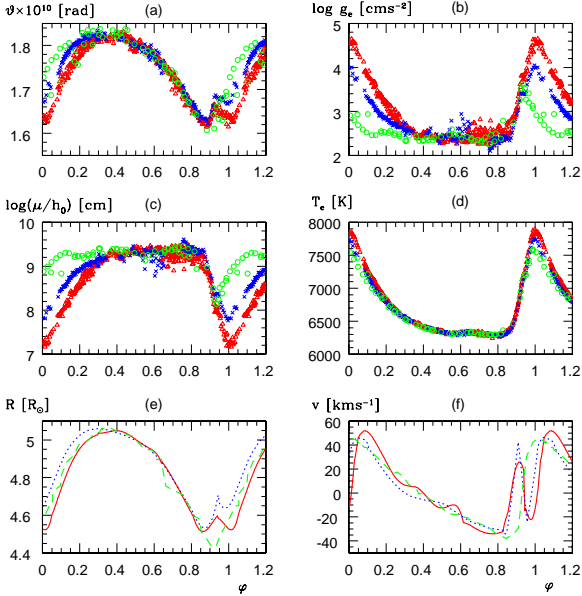}
\vspace{-0.8cm}
\caption{The variable parameters of SW And as a function of
 phase in absolute units.
  Panel (a): the angular radius
 $\vartheta(\varphi)=R(\varphi)d^{-1}$,  
 panel (b): the effective gravity $\log g_{\rm e}(\varphi)$,
 panel (c): the barometric scale height for unit averaged
            molecular mass
	    $\mu$ 
            at the top of the atmosphere,   
 $h_0(\varphi)=g_{\rm e}/{\cal R}T(R,\varphi)$,
 panel (d): effective temperature $T_{\rm e}(\varphi)$. 
 Panel (e): the radius of zero optical depth $R(\varphi)$,
 (green) dashed line: radius variation integrated from the radial
 velocity of \citet{liuj2},
 panel (f): the velocity $v(R,\varphi)$,
 the (green) dashed line  
 $v_{\rm puls}(\varphi)$ was computed from
 $v_{\rm rad}(\varphi)$ of \citet{liuj2} with
 ${\cal P}_p=1.32$, 
 $v_\gamma=-20.9\mbox{kms}^{-1}$ 
 in Eq. (\ref{5.101}).
 Green circles in panels (a)-(d) and dashed (green) lines in panels (e), (f): 
 folded from 
 $\rm{HJD}-2447100=20.57\mbox{-}{23.86}$
 \citep{liuj2}.
 Red triangles in panels (a)-(d) and (red) lines in panels (e), (f): 
 folded from
 ${\rm HJD}-2454300=50.4\mbox{-}53.8$.
 Blue crosses in panels (a)-(d) and dotted (blue) lines in panels (e), (f): 
 folded from
 ${\rm HJD}-2454800=22.2-32.37$. 
 }
\label{fig2}
\end{figure}

\begin{figure}
  \includegraphics[width=84mm]{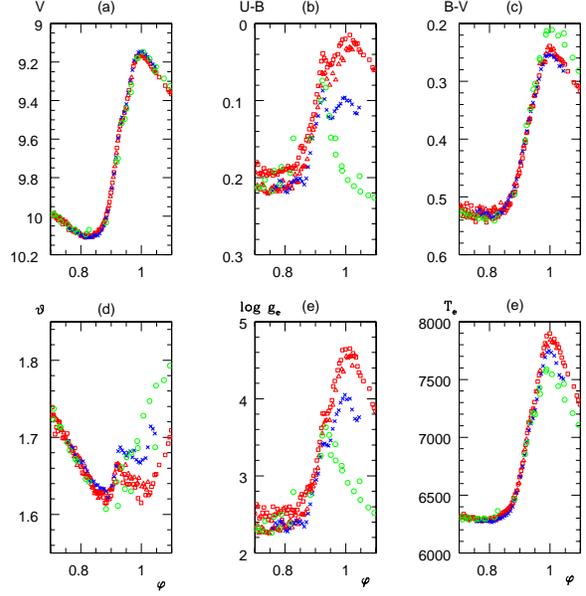}
\vspace{-0.8cm}
\caption{Zoomed characteristic parameters of SW And
  during the most compressed state of the atmosphere.
  Green circles: folded from $\rm{HJD}-2447100=20.57\mbox{-}{23.86}$
   \citep{liuj1},
  (red) triangles: folded from $\rm{HJD}-2454300=50.5991\mbox{-}50.7399$,
  (red) squares: folded from $\rm{HJD}-2454300=51.4821\mbox{-}51.7024$,
  (blue) crosses: folded from $\rm{HJD}-2454832=0.2465\mbox{-}0.3726$.
  }
\label{fig5a}
\end{figure}

Two particular features are worth of mentioning from
Fig.~\ref{fig2}. To show them clearly the zoomed variation of 
{\it V\/}, {\it U--B\/}, {\it B--V\/}, 
$\vartheta$, $\log g_{\rm e}$,
$T_{\rm e}$ are plotted in Fig.~\ref{fig5a} as a function of
$\varphi$. 
 
Two atmospheric shocks can be observed on the ascending
branch separated at
$\varphi\approx 0.95$.
The first one is in the interval
$0.8 \la \varphi \la 0.95$, this is the precursor shock
\citep{smith1}, it
is identical in each epoch, a rapid expansion of the atmosphere
starts.
The `main' shock, the occurrence of another rapid
expansion at
$0.95 \la \varphi \la 1.01$
is very different: it was missing, strong and medium strong 
during the observations at
$\mbox{HJD}-2444720=0\mbox{-}3$,
$\mbox{HJD}-2454350=0\mbox{-}3$, 
$\mbox{HJD}-2454820=2\mbox{-}12$,
respectively.  
This double structure of the shock is reflected by
the rapid change of the velocity 
$v(R,\varphi)$
in the interval
$0.9\la\varphi\la 1.1$:
the line and the dotted line in Fig.~\ref{fig2}f have two 
almost identical maxima separated by the minimal
$v(R,\varphi\approx 0.98)\approx -20\:\mbox{kms}^{-1}$,
that is by a short contraction episode.
The violation of QSAA is stronger in the main
shock than in the first one.

In addition to the maxima of
$v(R,\varphi)$
at
$\varphi\approx 0.9, \; 1.06$, 
two weak humps are visible at
$\varphi\approx 0.38, 0.55$
in the velocity curves from
$\rm{HJD} -2454350=0,1,3$,
the hump at
$\varphi\approx 0.38$
is not visible on
$\rm{HJD}\approx 2454800$.
The undulation at
$\vartheta\approx 0.55$
is visible in the material of \citealt{liuj1} as well. 
This latter might be the pre-precursor shock 
which was found in $\vartheta(\varphi)$
of the RRab star SU Dra (Fig.~1 in Paper I). 

It is obvious that the shock hitting the atmosphere 
is more of hydrodynamic than thermal nature: it is more 
pronounced in
$g_{\rm e}(\propto \varrho^{-1}\mbox{grad} p)$
than in
$T_{\rm e}$,
the increments are
about one dex and 4~per cent,
respectively, where 
$\varrho(r), p(r)$
are the density and pressure in the atmosphere.

\subsection{DH Peg}

The 24 shock-free phase points 
$0.1 <\varphi < 0.4$
on
${\rm HJD}=2454349$ 
were used in the minimization process,
an equivocal result was found in the interval
$0\le E(B-V)\le 0.1$,
therefore, we adopt
$E(B-V)=0.08$
\citep{rodn1} derived from the photometry in the Walraven system
\citep{lub1}.  
$E(B-V)=0.08$
yields
$[M/H]=-0.35\pm 0.1$~dex
from the minimization process, we use this value of metallicity.
The errors are
$\overline{\Delta T_{\rm e}}=17$~K and
$\overline{\Delta \log g_{\rm e}}=0.071$ 
in the minimum. This metallicity differs from
$[M/H]=-0.8$~dex
obtained from a differential
curve-of-growth analysis of DH Peg \citep{butle1}, however, it is
in good agreement with
$[M/H]=-0.42$~dex
derived from the Preston-index \citep{butle1}.
(We remark that adopting
$[M/H]=-0.8$~dex 
would result in the increase to
$\overline{\Delta T_{\rm e}}=24$~K
and
$\overline{\Delta \log g_{\rm e}}=0.092$.) 

\begin{figure}
  \includegraphics[width=84mm]{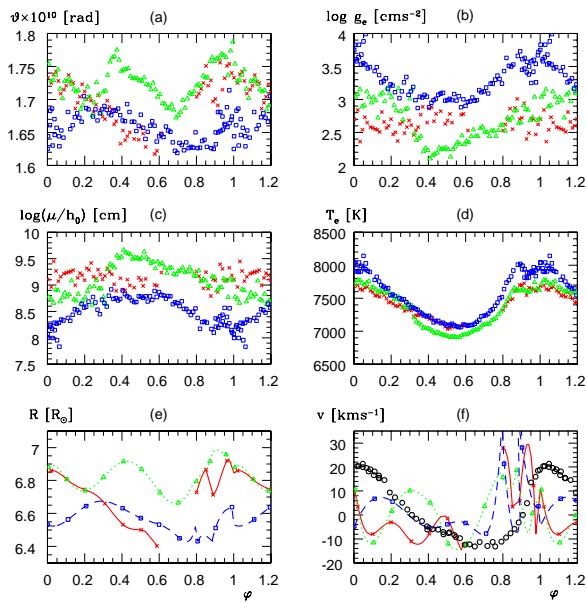}
\vspace{-0.8cm}
\caption{The variable parameters of DH Peg in absolute units.
 For the order of the panels see Fig.~\ref{fig2}.
 Red crosses, (green) triangles, (blue) squares:
 data from $\rm{HJD}-2454300=49,52,54$, respectively.
 Red line, (green) dotted, (blue) dashed in panels (e,f): 
 ${\rm HJD}-2454300=49, 52, 54$, respectively.
 Black circles: pulsation velocity from \citet{rodn1}}
\label{fig11}
\end{figure}

A preliminary analysis of the data
by the {\sc bbk} package revealed that the atmosphere of DH Peg
was almost continuously in a shocked state in the time intervals
HJD$-$2454300=52.4426-52.6952, 54.3792-54.6752
as it is obvious from Fig.~\ref{fig11} and from the averages
$\langle \log g_{\rm e} \rangle= 2.565, 2.640, 3.297$,
and
$\langle T_{\rm e} \rangle= 7337, 7323, 7554$~K
on
${\rm HJD}-2454300=49, 52, 54$,
respectively. Thus, we could use the 57 epochs of
$\rm{HJD}-2454349=0.6244-0.7057$
(phase intervals 
$0.0016 < \varphi < 0.5879$
and
$0.8005 < \varphi < 1.0$)
in the {\sc bbk} package to determine 
the data in Table \ref{tab5}.
The inclusion of the omitted epochs from
${\rm HJD}-2454300=52,54$
would increase 
$d$,
${\cal M}_{\rm a}$, 
$L_{\rm eq}$,
$R$,
shifting DH Peg in a range of these parameters which is
characteristic rather for a Cepheid or semi-regular variable. 

Remarkable peculiarities are revealed by Fig.~\ref{fig11}.
\begin{itemize}
\item[$\bullet$]
The minimal and maximal radius are at
$\varphi\approx 0.7,0.95$,
respectively, that is, too early with respect to 
the maximal
$\log g_{\rm e},T_{\rm e}$.
The phase dependence of the atmospheric kinematics of 
DH Peg is different in comparison to
an RRab star, e.g. SW And.  
\item[$\bullet$]
The amplitude of
$\log g_{\rm e}$
is small ($< 0.6$) over a time-scale
$0\fd26$ ($\approx P$).
This is characteristic for high amplitude 
$\delta$~Sct (HADS) or SX Phe stars. The
overall amplitude in the whole interval
${\rm HJD}-2454300=[49,54]$
is
$\approx 1.5$,
but even this higher value is below the characteristic value
($\ga 2$)
for a normal RR star. 
\item[$\bullet$]
On the other hand,
the average 
$\overline{\log g_{\rm e}(\varphi)}=2.86$,
$T_{\rm eq}=7464$~K
and
$g_{\rm s}\approx 4.5\:\mbox{ms}^{-2}$
place DH Peg among the RRc stars.
\end{itemize}
 
\subsection{CU Com}

$[M/H]=-2.2$ dex
for CU Com supports the appropriateness of our photometric
method since it is identical with the value
found by \citet{clem1} from high-dispersion
spectroscopy. 

\begin{figure}
  \includegraphics[width=84mm]{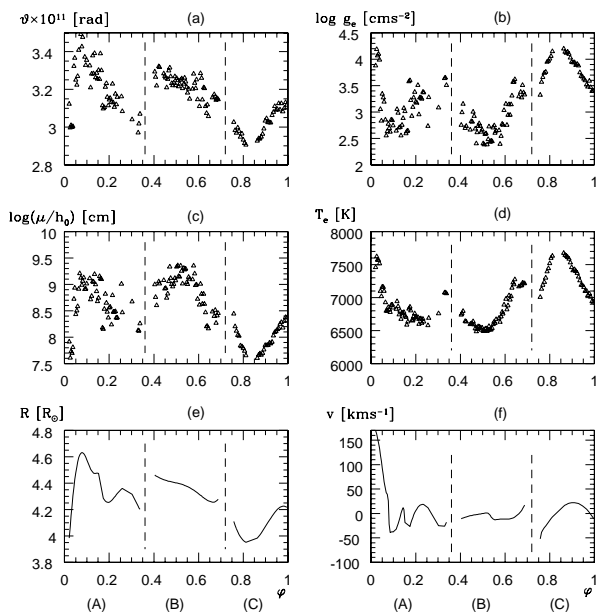}
\vspace{-0.8cm}
\caption{The variable parameters of CU Com at the light
 curve segments plotted in Fig.~\ref{fig5} in absolute units.
  For the order of the Panels see Fig.~\ref{fig2}.
  }
\label{fig10}
\end{figure}

The variable atmospheric parameters of CU Com are very similar
to those of an RRab star, the response of the atmosphere to
the pulsation does not differ from that of an RRab star. 
The presence of
the period of the first overtone with comparable amplitude
of the fundamental mode produces only a more complicated variability
of the motions because the maxima, minima of
the governing factors
$g_{\rm e}$,
$T_{\rm e}$
belonging to the two frequencies
are sometimes added or attenuated.
CU Com follows the rule of the RR stars mentioned in 
Sec.~\ref{sec4.4}:
it can be seen from Figs.~\ref{fig5} and \ref{fig10} 
that minimal radius, barometric scale height, maximal effective
gravity and temperature coincide with the minimal
{\it V\/},
{\it U--B\/},
etc.
There exist time intervals -- e.g. section (B) in Fig.~\ref{fig10}
-- when the atmosphere is shock-free, QSAA is a good approximation
and the dynamical corrections are small:
$a^{\rm(dyn)} \ll g_{\rm s}$.
The amplitude of
$\log g_{\rm e}$
is
$\ga 2$
which is the characteristic value for an RRab star.

\subsection{DY Peg}

$E(B-V)=0.0$
and
$[M/H]=-0.05$~dex
were found, this almost solar metallicity suggests 
that DY Peg is rather a HADS, not an SX Phe star, as it was 
classified on the basis 
$[M/H]=-0.7$~dex
\citep{andre1}.

\begin{figure}
  \includegraphics[width=84mm]{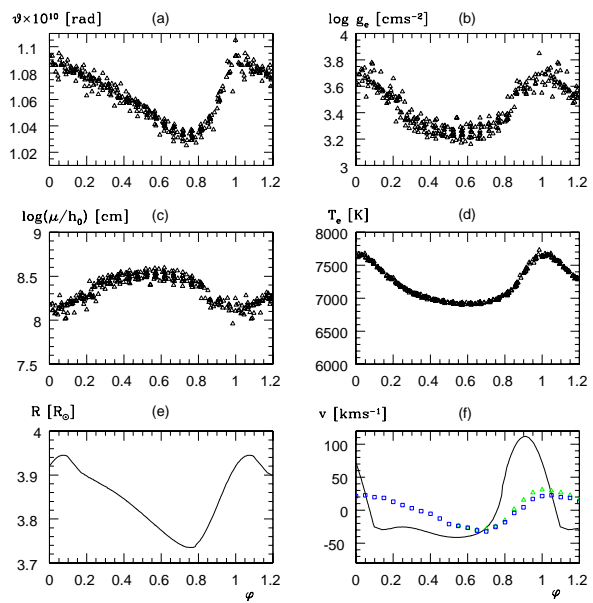}
\vspace{-0.8cm}
\caption{The variable parameters of DY Peg in absolute units.
 For the order of the panels see Fig.~\ref{fig2}.
 Green triangles and blue squares in panel (f): 
 $v_{\rm puls}(\varphi)$
 computed from the
 radial velocity curve of \citet{meyla1} and \citet{wilson1}
 by Eq. (\ref{5.101}) with
 ${\cal P}_p=1.41$,
 $v_\gamma=-25\mbox{kms}^{-1}$. 
 }
\label{fig9}
\end{figure}

The following pattern can be seen from Figs.~\ref{fig4} and
\ref{fig9}: maximal
$R$,
$g_{\rm e}$,
$T_{\rm e}$,
$\vartheta$
and minimal 
$h_0^{-1}$ 
appear simultaneously at the maximal brightness in 
{\it V\/}
which is accompanied, paradoxically, with maximal 
{\it U--B\/}.
This behaviour is different from the phase dependence at RRab 
stars, it is similar to that found at DH Peg. The average 
$\log g_{\rm e}=3.41$
is larger, its amplitude is small 
($\approx 0.6$)
compared to the typical values of an RRab star. This behaviour 
is explained qualitatively by the larger static
$g_{\rm s}=$30.2~ms$^{-2}$ 
which is an attenuating factor during the pulsation.

\section{Discussion}

As mentioned in Sec.~\ref{sec3}, the zero points
from the attempt of the tie-in observations  
had to be shifted in order to bring our
colours in coincidence with those of 
\citet{tifft1}, \citet{paczy1}, \citet{liuj2} and \citet{oja1}.
The uncertainty of zero points of magnitude scales, the only
dim term in our photometry, could not produce the observed
variation in the colour curves of SW And, DH Peg and
DY Peg. The study of the colour curves, especially
{\it U--B\/}
revealed that changes of the light curve having a 
small amplitude in
{\it V\/}
can be easier detected by extending observations in 
the ultraviolet colour.

The extensive multicolour observations including the 
{\it U\/}
band and using the different pairs
of colour indices of the
{\it UBV(RI)$_{\mathrm C}$\/}
photometry allow us to determine the main governing
atmospheric parameters 
$T_{\rm e}(\varphi)$,
$g_{\rm e}(\varphi)$
better than the hitherto applied methods based on a single
colour index using one (more or less arbitrarily 
chosen) colour index for
$T_{\rm e}(\varphi)$
and another one for
$g_{\rm e}(\varphi)$
(e. g. \citealt{rodn1,liuj2}).
Our method enables us to obtain a more thorough and
consistent information on the dynamical changes in the atmosphere 
of stars pulsating in radial modes. The
parameters 
${\cal M}_{\rm a}$
and the distance 
$d$
to the star are parameters in the hydrodynamic Euler
equation for the pulsation of the stellar atmosphere. 
(This perception allows us to determine 
the parameters on the basis of an astrophysical background.
The Euler equation is written in Euler formalism \citep{pring1}).
This is a dynamical method, it is completely different in
comparison with the Baade-Wesselink (BW) method which
is essentially a kinematic method, yielding 
$d$
as a main result. Other astrophysical methods, 
the theories of stellar evolution and pulsation, give masses
${\cal M}_{\rm ev}$, 
${\cal M}_{\rm puls}$
and luminosity, where 
$d$ 
is a derived quantity
from comparing the theoretical and observed luminosities.   

The accuracies are 10 and 2-3 percents for 
${\cal M}_{\rm a}d^{-2}$
and
$\vartheta(\varphi)$,
respectively, the differentiation of
$\vartheta$
with respect to time allows the determination of the angular 
velocity and the angular
acceleration of the pulsating atmosphere in the reference frame 
of the observer. They are used in the hydrodynamic equation
of motion in the stellar reference frame and permit
to clear up the
kinematics, dynamics of the pulsating atmosphere in more 
details than it could be done if the uniform atmosphere
approximation (UAA) was used (which was defined
in Paper I and is used also in any BW analysis.) 

Of course, the QSAA is assumed during the
whole cycle of the  pulsation. It can surely be regarded as a first
approximation, however, considerable corrections to QSAA can 
be expected
only in the shocked phases. They are beyond 
the scope of the present series of papers and they 
have negligible effect on the derived fundamental parameters
because
${\cal M}_{\rm a}d^{-2}$,
$d$
are determined from inverting the photometry in the shock-free phases,
and the other key quantity 
$\vartheta$ 
is also taken from the shock-free phases.

Now we discuss the results, 
compare them with the parameters 
obtained from a BW analysis and remarks are given on the
results for the stars. 

\subsection{Comparison of $d$ with trigonometric parallax data}

An important test for the reliability 
of our new photometric-hydrodynamic method is offered
if trigonometric parallax of the target stars
is available.

Parallax
$\pi=1.42\pm 0.16$~mas
was measured for SU Dra with
the Fine Guidance Sensor of the {\it Hubble Space Telescope}
\citep{bene1} yielding
$d=704\pm 79$~pc.
This value is in perfect agreement with 
$d=663\pm 67$~pc
(Paper I) and
$d=640\pm 77$~pc from a BW analysis \citep{liuj2}.

The satellite {\it Hipparcos} \citep{perry1} measured
$\pi=-0.04\pm 1.50,\; 0.15\pm 1.42,\; 0.36\pm 2.02,\; 
1.11\pm 1.15$~mas
for SW And, DH Peg, DY Peg and SU Dra, respectively. 
The revised values are
$\pi=1.48\pm 1.21,\; -2.89\pm 1.71,\; -1.22\pm 1.60,\; 
0.20\pm 1.13$~mas
\citep{tycho2}.
These values can be regarded 
as a null result. However, 
$d=470\pm 40$~pc
of DH Peg \citep{rodn1} and 
$d=250\pm 40$~pc
of DY Peg \citep{burki1} from a BW analysis must probably be
too small because they yield
$\pi=2.1$ and 4.0~mas
for these stars. Parallaxes of these values could have been measured
by {\it Hipparcos}.

\subsection{Comparison of $v(R,\varphi)$ with pulsation
velocity derived from radial velocity observations}

If the radial velocity 
$v_{\rm rad}(\varphi)$
is observed and the equation of type
\begin{equation}\label{5.100}
v(r,\varphi)={\hat{\cal P}}
      \{v_{\rm rad}(\varphi)-v_\gamma\}
\end{equation}       
can be solved, it is possible to determine
$d$
because the depth dependent pulsation velocity 
$v(r,\varphi)$
is known from our method and it depends on
$d$,
e.g. in a form given in Paper I,  
$\hat{\cal P}$,
and
$v_\gamma$
are a projection operator and
the centre-of-mass velocity of the star, respectively. 
The velocity of the zero optical depth is
$v_{\rm puls}(\varphi)=v(R,\varphi)$.

The simplified form of solving Eq. (\ref{5.100}),
\begin{equation}\label{5.101}
v_{\rm puls}(\varphi)
    =-{\cal P}_p[v_{\rm rad}(\varphi)
       -v_\gamma], 
\end{equation}
is used in a BW analysis (\citealt{liuj1, rodn2} etc.). 
(${\cal P}_p=\textstyle{{24}\over{17}}$ or $\textstyle{{3}\over{2}}$
for a spectral line of infinitesimal width in an atmosphere
if the velocity is depth-independent and limb darkening of the
velocity is or is not taken into account \citep{getting}.)
In actual applications, 
\begin{itemize}
\item[$\bullet$] a constant
($1.3 \la {\cal P}_p \la 1.4$) or even a
$\varphi$-dependent
${\cal P}_p$
are assumed \citep{mare1}.
\item[$\bullet$]
$v_{\rm rad}(\varphi)$
is taken from a photometric correlation of template spectra 
of non-variable stars with 
the spectra of the RR star having a variable spectrum, 
\item[$\bullet$]
$v_{\rm puls}=\dot\vartheta d$
is assumed, 
\item[$\bullet$]
$v_\gamma$
is obtained from integrating
$v_{\rm rad}(\varphi)$
and equating the upward and downward motions. 
\end{itemize}
The radial velocity is derived from
the masking technique (e.g. CORAVEL \citealt{liuj1})
or high dispersion
spectra in a limited interval of wavelength (e.g. \citealt{rodn2}).
It is integral of a depth-dependent velocity field
where the weight function is not known.
We note that neglecting the velocity gradient in the
pulsating atmosphere, that is, taking
$v(R,\varphi)={\dot\vartheta}d$,
and assuming a constant
${\cal P}_p$ 
results in UAA. It is a first approximation which ought to be 
refined. This refinement has, however, never been discussed
in the numerous papers dealing with the BW method, despite
of the indication of the considerable velocity gradient in
the pulsating atmosphere (see e.g. \citealt*{oke1}). 

In addition to the problem of the velocity gradient in
a pulsating, compressible stellar atmosphere,
a practical problem of the BW study 
is the strong dependence of the error
$\Delta d/d$
on the error
$\Delta v_\gamma$
discussed in \citet{gaut1} and Paper I. 
To determine
$v_\gamma$
a way consistent with Eq. (\ref{5.101})
would be to select the phases 
$\varphi_0$
when
$v_{\rm puls}(\varphi_0)=0$
and then
$v_{\rm rad}(\varphi_0)=v_\gamma$.
This is, however, not free of problems as it can be demonstrated 
by looking on panel (f) of Figs.~\ref{fig2}, \ref{fig11}, and
\ref{fig9}, because
$v(R,\varphi)$
has a complex structure and there exist a considerable
phase lag between
$v_{\rm puls}(\varphi)$
by Eq. (\ref{5.101}) and
$v(R,\varphi)={\dot\vartheta}(\varphi)d+\cdots$,
that is
$v_{\rm puls}(\varphi)$
and
$v(R,\varphi)$
are not comparable quantities. Smoothing the curves, adding
an arbitrary phase lag 
$v_{\rm rad}(\varphi+\delta\varphi)$,
$\delta\varphi\not=0$
and neglecting the additional terms of
$v(R,\varphi)$ 
can formally solve the
problem, however, it is cannot be motivated astrophysically. 
Qualitatively, it is obvious that the phase lag between the
two curves is a consequence of the neglected velocity gradient 
in the expanding-contracting atmosphere.

The above problem is only hidden but not solved if
the radius displacement
\begin{equation}\label{5.102}
\Delta R=\int_{\varphi(t_1)}^{\varphi(t_2)}
v_{\rm puls}[\varphi(t)+\delta\varphi]{\rm d}t
\end{equation}
is compared with
$\Delta\vartheta d={\vartheta(t)}\vert_{t_2}^{t_1}d$
in the frame of a BW analysis.

The details of the problem can be visualized by the 
curves of SW And (Fig.~\ref{fig2}e,f.)
The value of the phase lag
$\delta\varphi$
depends upon whether 

\noindent
$R(\varphi)$
and
$\langle R \rangle +\Delta R$
or 

\noindent
$v(R,\varphi\approx 0.92)$
and
$v_{\rm puls}(\varphi)$
or

\noindent
$v(R,\varphi\approx 1.05)$
and
$v_{\rm puls}(\varphi)$

\noindent
are brought to coincidence.
The radius displacements 
$\langle R\rangle+\Delta R(\varphi)$
calculated from Eq. (\ref{5.102}) are
plotted in Fig.~\ref{fig2}e with
$\langle R\rangle=4.78$~R$_\odot$
and 
$\delta\varphi=0$.
The double peak of
$v(R,\varphi)$,
is not visible in 
$v_{\rm puls}(\varphi)$
at all, perhaps because of the scarce sampling of
$v_{\rm rad}(\varphi)$
(Fig.~\ref{fig2}f).
The panels demonstrate that there is an 
uncertain phase lag 
($\vert\delta\varphi\vert\la 0.1$).
The fine structure of
$R(\varphi)$
cannot be reproduced by the integration of the radial velocity
if Eqs. (\ref{5.101}) and (\ref{5.102}) are used. Reliable 
$R(\varphi)$
could only be obtained if a better solution of Eq. (\ref{5.100}) 
were applied than Eq. (\ref{5.101}).

The motion of the atmosphere is derived in this series of
papers from
$d, {\cal M}_{\rm a}$, 
$\vartheta(t),T_{\rm e}(t),\log g_{\rm e}(t)$
in absolute units. It is compared with the observed
radial velocity data of SW And, DH Peg and
DY Peg in Panels (f) of Figs.~\ref{fig2}, \ref{fig11}, and
\ref{fig9}. 
The difference of
$v_{\rm puls}(\varphi)$
and
$v(R,\varphi)$ 
can partially be explained by the simplifications involved in a
BW analysis and, additionally,  
by the difference of the characteristic
time of a multicolour observation ($\la 3$~min) and the
longer exposure time to take a spectrum. The scanty
sampling has a smoothing effect, e. g. the
complex structure of
$R(\varphi)$
of DH Peg cannot be explored by numerically integrating 
$v_{\rm puls}(t)$
sampled in intervals 10-30 min. 

These considerations substantiate why we trust better 
in the distances from the present study
if there is a significant
difference between them and those from a BW analysis. 

\subsection{Remarks on the observed stars}

\begin{figure}
  \includegraphics[width=84mm]{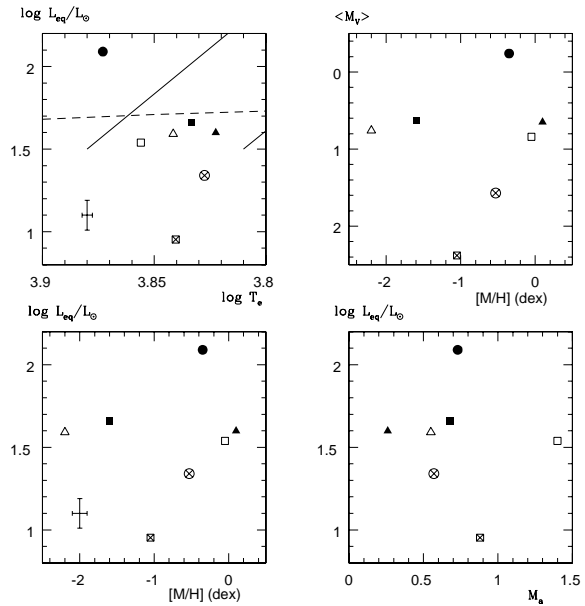}
\vspace{-0.8cm}
\caption{Luminosity - effective temperature, 
	 de-reddened visual absolute magnitude - metallicity, 
         luminosity - metallicity and
	 mass - luminosity diagrams of the observed
	 stars. 
	 Filled circle: DH Peg, 
	 square: DY Peg, 
	 triangle: CU Com, 
	 filled square: SU Dra,
	 filled triangle: SW And,
	 circle with cross: V372 Ser, 
	 square with cross: V500 Hya. 
	 The estimated errors are
	 plotted in the lower left part of the left panels.
	 Lines in the upper left panel: 
	 blue and red edge of the instability strip, 
	 dashed line: zero-age horizontal branch of M3
	 \citep{agui1}    
 }
\label{fig10a}
\end{figure}

Fig.~\ref{fig10a} is a plot of some characteristic 
results given in Tables~\ref{tab4}, \ref{tab5}. It gives an
impression on the diversity of the physical parameters
of the target stars which form a more or less uniform group of
radially pulsating field stars if they are classified by 
periods and amplitudes in one broad optical band.

The upper left panel in Fig.~\ref{fig10a} is a plot of
$\log T_{\rm eq}$ vs. $\log L_{\rm eq}$,
a Hertzsprung-Russell diagram in terms of absolute 
astrophysical units showing 
a part of the instability strip.
It is obvious that SW And, SU Dra, CU Com, 
DY Peg form a group (regular RR stars), while 
DH Peg is well above the
horizontal branch formed by this group. V500 Hya is
at the position expected for a SX Phe star, V372
Ser is at half way between SX Phe and RR stars. 
This region was not explored by
detailed hydrodynamic calculations, however, Fig.~3 of
\citealt{szab1} indicates the possibility of stable DM
pulsation towards lower luminosities.

Using the regular stars only the following relations
can be derived for a dependence on metallicity.
\begin{eqnarray}
\label{6.100}
\log L_{\rm eq}&=&(-0.02152\times[M/H]+1.5775 \pm 0.0538)\mathrm{L}_\odot \\
\label{6.101}
\langle M_V \rangle&=&0.01103\times[M/H]+0.7303 \pm 0.1194,
\end{eqnarray}
The relations have actually a null slope. Furthermore, the
independence of luminosity on 
${\cal M}_{\rm a}$
is remarkable.

\subsubsection{SW And}

Significant differences have been found in the photometric
and physical parameters on
$\mbox{HJD}-2444720=0\mbox{-}3$,
$\mbox{HJD}-2454350=0\mbox{-}3$ and
$\mbox{HJD}-2454820=2\mbox{-}12$:
$L_{\rm eq}=41.8$, 39.8 and 41.5~L$_\odot$;
$T_{\rm eq}=6690$, 6644 and 6672~K,
$\langle M_V \rangle=0.66$, 0.71 and 0.68~mag
(see Tables~\ref{tab3}, \ref{tab4}, \ref{tab5}).
These variations confirm that 
$P$
in Table~\ref{tab1} is not the sole period in the 
atmospheric pulsation. 

SW And follows the rule  
that minimal radius, barometric scale height and maximal
$T_{\rm e}$,
$g_{\rm e}$
belong to maximal brightness in 
{\it V\/}
and minimal
{\it U--B\/} 
(Figs.~\ref{fig1} and \ref{fig2}), this is common in RRab stars.

The derived very low mass 
${\cal M}_{\rm a}=0.26$~M$_\odot$
is surprising. It
was obtained from a
background provided by theoretical model atmospheres and
some hydrodynamics, not using the theory of
pulsation and evolution of RR stars \citep{smith1}.
It is much lower than the canonical value 
$\approx 0.6$~M$_\odot$
of an RRab star \citep{liuj2, rodn2}
or
$\approx 0.5$~M$_\odot$
from the empirical Fourier fitting technique for metal rich 
RR stars \citep{neme2}. 
It is a serious challenge to the theories of stellar pulsation
and evolution. 

We remark that 
the mass of an RR Lyrae-like star in the eclipsing binary system
OGLE-BLG-RRLYR-02792 was found previously to be
$0.26\pm 0.015$~M$_\odot$
\citep{pietrz1} and has been
substantiated by some theoretical considerations \citep{smole1}. 
This favours the adoption of our anomalous mass value.
Common features of SW And with OGLE-BLG-RRLYR-02792 are  
the bump in the middle of the ascending branch, and the similar
luminosity and effective temperature
$L_{\rm eq}\approx 39$~L$_\odot$, 
$T_{\rm eq}\approx 6664$~K.
(\citealt{smole1} found them as: $L_{\rm eq}\approx 33$~L$_\odot$ and
$T_{\rm eq}\approx 6970$~K.)

The distances
$d=511\pm 56,\; 481\pm 33$~pc
were adopted by \citet{liuj2} and \citet{rodn2}, respectively,
these are some 20 per cent smaller than our
$d=626\pm 31$~pc.
The main argument to accept our larger distance to SW And is
that
${\cal M}_{\rm a}d^{-2}=(7.00\pm 0.76)\times 10^{-7}$
(Table~\ref{tab5}) was determined from the shock-free state
of the atmosphere when the QSAA is expected to be a
reliable approximation on an astrophysical basis. 
The value of
${\cal M}_{\rm a}d^{-2}$
is fairly firm central point in our theory. An
adoption of the distances 
$d=511, 481$~pc
would reduce the mass
${\cal M}_{\rm a}$
to a more anomalous low value
$\la 0.17\mbox{M}_\odot$. 

The difference between
$d=511$~pc
and 
$481$~pc
can be attributed to the difference between
$v_\gamma=-20.9$
and
$-19.2$~kms$^{-1}$
in \citet{liuj2} and \citet{rodn2}, respectively, because an error
1~kms$^{-1}$
of 
$v_\gamma$
results in an error
$\Delta d / d\approx 0.1$
(\citealt{gaut1}, Paper I).
The uncertainty of 
$v_\gamma$
is obvious: 
$v_{\rm puls}(\varphi)\approx 0$ 
at
$\varphi=0.4\pm 0.01$ and $0.3\pm 0.01$
from the IAC80 and RCC observations, respectively. The
interpolated centre-of-mass velocities are
$v_{\rm rad}=-21\pm 1$, $-28\pm 1$~kms$^{-1}$
($\approx v_\gamma$ !)
from the observations of \citet{liuj2}. The change of  
$v_\gamma$
from
$-21$
to
$-28$~kms$^{-1}$
within 
480~days
might even be an indication for the binarity of SW And
supported by the anomalous 
${\cal M}_{\rm a}=0.26$~M$_\odot$. Or more likely, it is an 
artifact yielded by the use of Eq. (\ref{5.101}).

The differences in 
$\langle R \rangle=4.36,4.16$~R$_\odot$
(\citealt{liuj2}, \citealt{rodn2})
and
$4.78$~R$_\odot$
(Table~\ref{tab5}), respectively, originate mainly from the
different values of 
$d$.
The derived photometric angular diameters of SW And 
do not differ significantly in the three studies. 

\subsubsection{DH Peg}

The instability in
({\it U--B})$(\varphi)$
of DH Peg is known from previous observations and this seems 
to be a common feature 
of RRc stars \citep{tifft1}. Our observations 
confirm the instability.   
The presence of other frequencies in RRc stars \citep{moska1} with 
increasing amplitude toward the ultraviolet
could be a natural explanation of the non-repetitive character
of the colour curves toward the ultraviolet and of the
larger scatter in the descending branch of the near infrared
colour curves. Like at SW And, the observation
and analysis of the ultraviolet part of the spectrum have lead
to a more detailed picture on the pulsation properties.
The time span of our observations is
short to draw a more definite conclusion.
 
The ambiguity  
$[M/H]=-0.35\: \mbox{or} \: -0.8$~dex 
for DH Peg could have been caused
by the long term variation of the light curve and 
the break down of QSAA for the rapidly and intensively
changing atmospheric conditions. A binning of 
non-coherent zero points in our differential photometry
seems to be an unreal possibility. Eventual adoption of
$[M/H]=-0.8$~dex
would not result in a considerable change of the anomalous
parameters in Table~\ref{tab5}, but the decrease of
$30$ 
to 
$21$ 
pairs of colour indices giving a solution for
$T_{\rm e}$
and
$\log g_{\rm e}$
speaks strongly against 
$[M/H]=-0.8$~dex.

Over a time interval of a few pulsation cycles,
the large and systematic difference of the physical parameters,
that is, approximately one dex difference in
$\log g_{\rm e}$
and barometric scale height
$h_0^{-1}$,
the complex structure of 
$R(\varphi)$
and 
$v(R,\varphi)$  
suggest that periodic or stochastic variations are present 
in the atmosphere in addition to the period
$P$
in Table \ref{tab1}.

Furthermore, the violation of the assumptions involved in the QSAA
and eventual excitation of non-radial mode(s) with large
amplitude can also be a distorting factor for the derived
fundamental parameters because our simplified 
(spherically symmetric) hydrodynamic
model is not able to describe this type of the atmospheric
pulsation.
The large value of the dynamical correction term
($\vert a^{\rm (dyn)} \vert \ga g_{\rm e}$)
during the whole observed phases on 
${\rm HJD}=2454354$ 
substantiates this conjecture, e.g.
$\langle\log g_{\rm e}\rangle\approx 3.3$
was during this period, some
$0.6$~dex higher than on
${\rm HJD}=2454349$.
Large deviation from spherical symmetry would modify the colours 
compared to those of the ATLAS models. 

Some cautiousness is appropriate in connection with the
fundamental parameters of DH Peg derived by \citet{rodn1} 
from BW analysis. Their distance, consequently
their luminosity and
$\langle R\rangle$ differ significantly from the values in
Table~\ref{tab5}. 
The upper and lowers limit of the photometric angular radius
are in good agreement with our ones,
however, there is a systematic difference in their  
$\vartheta(\varphi)$
derived from the different colour combinations and the
cycle to cycle variations over a few pulsation period
are ignored. Finally, they select
one colour index to determine
$\vartheta(\varphi)$ 
without an attempt to reconcile
the values from the different colour indices by taking into 
account the effect of the variable
$\log g_{\rm e}(\varphi)$. 
The upper and lower limits originating from two different
colour indices 
$6800\:\mbox{K} < T_{\rm e}(\varphi) < 7800\:\mbox{K}$ are
in coincidence with our values plotted in Fig. \ref{fig11}.
However, the problem of selecting one colour index reducing
the upper or increasing the above lower limits
is again not solved.

The \citealt{rodn1}
$470\pm 40$~pc
distance could reduce the luminosity to
34.1~L$_\odot$, 
the minimal and maximal radius to 
[3.37,3.72]~R$_\odot$.
However, the complexity and the amplitude of 
$v(R,\varphi)$
cannot be reconciled with the rather smooth
$v_{\rm puls}(\varphi)$
(circles
in Fig. \ref{fig11}e) derived from observed radial velocities
by Eq. (\ref{5.101}). 
Furthermore, a large phase lag 
($\Delta\varphi\approx -0.17$)
is necessary to bring the maxima of the
two curves in coincidence. The null result of the {\it Hipparcos}
parallax suggests larger 
$d$. 

The following dilemma has emerged from the results. DH 
Peg is either an anomalous RR star with anomalous luminosity 
$L_{\rm eq}\approx 130$~L$_\odot$
and radius 
$R\approx 7$~R$_\odot$
exceeding the
canonical values \citep{smith1}, or it is a low mass Cepheid
with  
$g_{\rm s}, T_{\rm e}$
exceeding the canonical values
$0.01 \la g_{\rm s} \la 1\:\mbox{ms}^{-2}$,
$T_{\rm e}\approx 5600$~K
\citep{mare1}. 
Some cautiousness is
appropriate concerning the row DH Peg in Table \ref{tab5},
however, our numerous attempts to revise 
them closer to the canonical values of RR stars could not
lead to more conventional values. We think that a natural 
resolution of the dilemma is that the conditions
${\rm C^{(I)}}$
and
${\rm C^{(II)}}$
of QSAA are not satisfied in DH Peg.

\subsubsection{CU Com}

CU Com has a Galactic height 
$\approx 3$~kpc
and has a very low metallicity
$[M/H]=-2.2$ dex.
These data show that CU Com might be a member of the outer
halo population \citep{natur1}. Its mass, radius,
$L_{\rm eq}$,
$T_{\rm eq}$,
$g_{\rm s}$
agree well with the canonical values for RRd stars. 
The mass
$0.55\pm 0.03\; \mbox{M}_\odot$
(Table~\ref{tab5}) differs from the pulsation mass 
$0.830\pm 0.005$
derived from a (several times revised) Petersen diagram
\citep{clem1}.

\subsubsection{DY Peg}

The conditions
${\rm C^{(I)}}$
and
${\rm C^{(II)}}$
(Paper I) are satisfied in all phases of DY Peg. The rapid 
change of
the atmospheric parameters during the very short period was
not found to be a hindrance to determine the
astrophysical parameters. This demonstrates that
our method is robust and works well for all types of radially
pulsating stars with large amplitude.

The scatter of
$\vartheta(\varphi)$,
$\log g_{\rm e}(\varphi)$,
$h_0(\varphi)$,
$T_{\rm e}(\varphi)$
is a consequence of folding the photometry with the main period
$P=0\fd072926492$. The noise from probable other frequency 
mentioned in Sec.~\ref{sec4.4} and the non-photometric
quality of the sky cannot not be separated by folding. 

The distance
$d=250$~pc
from the BW analysis \citep{burki1} would yield
$L_{\rm eq}\approx 3.2$~L$_\odot$,
$\langle R \rangle=1.17$~R$_\odot$,
${\cal M}_{\rm a}=2.42\times 10^{-7}\times 250^2
                 =0.13$~M$_\odot$.
These values are in contradiction with the present 
knowledge on HADS or SX Phe stars. A distance below
$500$~pc can be excluded on this basis. 
Application of an arbitrary phase lag 
$\Delta\varphi=-0.12$
in Fig.~\ref{fig9}f could result in a better agreement of 
$v(R,\varphi)$
(calculated as given in Paper II) and
$v_{\rm puls}(\varphi)$
calculated from the 
radial velocities \citep{meyla1,wils1} by Eq. (\ref{5.101}). 
The presence of the phase lag and the uncertain value of
$v_{\gamma}$
might be responsible for the very small
$d$
obtained from the BW analysis.

${\cal M}_{\rm a}=1.40{\cal M}_\odot$,
$g_{\rm s}=30.2$~ms$^{-2}$,
$[M/H]=-0.05$~dex
are characteristic for a HADS star.
It is a paradoxical situation that
$L_{\rm eq}=34.6$~L$_\odot$,
$\langle R \rangle=3.85$~R$_\odot$
and the position in Fig.~\ref{fig10a}
place DY Peg among the RR 
stars, but its short period 
$P=0\fd072926492$
is approximately one third of the period of the RRc stars
with shortest period. The short period is the strongest 
argument against being an RR star. 

\section{Conclusions}

The most comprehensive
$UBV(RI_C)$
photometry and its interpretation using ATLAS atmospheric
models have been reported in this series of papers
for the field stars 
SU Dra, V500 Hya, V372 Ser, SW And, DH Peg, CU Com, DY Peg. 
The results could be obtained only by simultaneous multicolour
observations, including an ultraviolet band is important! The 
interpretation has been done by 
simplified hydrodynamics using the mass and 
momentum conservation, that is: by the continuity and Euler
equations in Euler formalism
(in the spherical reference system of the star). 
These laws were applied first time in the study
of pulsating stars, their application belongs to the essence
of our method. This method is pioneer in
using a mixture of photometry and theory of stellar atmospheres.

The new method has rendered possible to determine
the mass, distance and the better
classification of radially pulsating stars. This could not
be done merely by analysing frequencies or properties,
parameters of light curves in a single (easily available)
broad optical band. As a by-product, the method offers a
better insight into the motion, dynamics of a 
pulsating stellar atmosphere. 

The results and the physical parameters of the stars
have been summarised in tabular form (Tables~\ref{tab3},
\ref{tab4}, \ref{tab5}) and snapshots have been given about
the variable atmospheric parameters.

\begin{itemize} 
\item[$\bullet$]
{\it Conclusions from the photometry.} The
$U$
observations, the colour curves
$U-B$ 
have revealed a new complexity
of the variability and
have rendered possible to discover or confirm the unstable
character of the light curve of SW And, DH Peg, and to a lesser
extent in DY Peg. These findings suggest that
either additional frequencies or some stochastic phenomena are
present in the atmospheric response to the
pulsation emerging from the sub-atmospheric layers. This result
is a challenge to further theoretical investigations
concerning pulsation and evolution theory.

\item[$\bullet$]
{\it Method of interpreting the photometry.} The 
$UBV(RI)_C$
observations were interpreted by the ATLAS atmospheric
models by comparing the theoretically calculated
colours and colour indices \citep{kuru1} with the observed ones.
QSAA was assumed in all phases. 
All colour indices were used. It has been demonstrated that
the use of one, more or less arbitrarily chosen, colour index for 
$T_{\rm e}$,
another one for
$g_{\rm e}$
results in information loss and systematic errors.  

\item[$\bullet$]
{\it Our photometric-hydrodynamic method} 
has been applied for all types of pulsating giant or sub-giant stars
with large amplitude, that is 
for RRab, RRc, RRd, SX Phe, HADS stars. 
The quantitative conditions (expounded in Paper I) 
have been applied to explore the phase intervals of
the pulsation when the assumptions of the QSAA are 
satisfied. The derived 
parameters represent a first approximation
in the phase intervals when the conditions of the QSAA are
violated 
because of the presence of shocks in the atmosphere. The
derivation of the corrections to
$g_{\rm e}(\varphi)$
and
$T_{\rm e}(\varphi)$,
consequently, to
$L_{\rm eq}$,
and
$T_{\rm eq}$
are beyond the scope of the
present series of papers. The (non-variable) fundamental 
parameters 
${\cal M}_{\rm a}$,
$d$,
$[M/H]$
were determined from the shock-free phases when QSAA
is expected to give the values of the phase dependent parameters
correctly.

\item[$\bullet$] {\it A comparison with the BW analysis.}
The essence of a BW analysis, 
the problem of comparing
$\vartheta(\varphi)$ 
with radius displacement derived from the integration of the radial
velocity curves 
has been discussed in some more details because the combination of
$\vartheta(\varphi)$, 
$T_{\rm e}(\varphi)$,
$g_{\rm e}(\varphi)$
reveal fine details of the atmospheric kinematics and
dynamics better. A better
description and understanding of the pulsating atmosphere is given
by our method. It has enabled us to derive the fundamental
parameters with less uncertainty.

\item[$\bullet$] 
{\it Remarks on the observed stars.}
Our method has yielded 
$d$
for the first time for DM stars 
because BW analysis has not been done for this type. The
fundamental parameters of CU Com
and V500 Hya have been found to be in the domain which is
characteristic for regular RRd and SX Phe stars, respectively. 

SW And has been found to be an RR star with very low mass
${\cal M}_{\rm a}=0.26$~M$_\odot$, 
furthermore, its ultraviolet light curve has been found to
be different over a time-scale of years.

DY Peg has been found to be a star with `mixed' parameters,
characteristic partially for RR stars, partially for HADS stars.

The known instability of the ultraviolet light curve of DH Peg
has been described in details from our observations.
We find that the assumptions of our
photometric-hydrodynamic 
method are probably not satisfied in this star.
Therefore, its determined fundamental parameters from our method,
$L_{\rm eq}\approx 130$~L$_\odot$,
$\langle R\rangle\approx 7$~R$_\odot$,
need a revision taking into account the dynamical corrections
with respect to the QSAA. 

V500 Hya and V372 Ser are RRd stars
if they are classified by frequencies. 
However, our results have shown that V500 Hya is an SX Phe
star with two frequencies of large amplitude. 
Our observations have pointed out an infrared excess 
$\approx 0.1$~mag for V500 Hya
which we could not interpret. It might indicate the presence
of a close companion.
A sub-luminosity ($\approx$ factor of 2) has been found for V372 Ser.

Our distance to SU Dra has been verified by
the {\it Hubble Space Telescope}. The
distance to SU Dra has been found to be in harmony with
that from the BW method. This latter is in spite of our reservation
concerning the BW method. This means that a good combination of
the uncertain points in a BW analysis can result in correct
fundamental parameters. However, the fundamental parameter, mass 
(${\cal M}_{\rm a}$)
originates from our method directly while it is a subject of
assumption in BW analysis from other theories like stellar
evolution and pulsation. 
\end{itemize} 

The results have shown that astrophysical study of radially
pulsating stars is not a boring theme.
 
\section*{Acknowledgements}

We are grateful for the travel support provided by the Hungarian 
Astronomical Foundation and for the hospitality at the Teide
Observatory, IAC. We thank J. M. Nemec, the referee of the
paper, and R. Szab\'o for comments and 
improving the English presentation.
We have used the SIMBAD data basis of CDS.

\bsp

\label{lastpage}

\end{document}